\documentclass[bibyear]{aa}
\usepackage{graphicx}
\usepackage{txfonts}
\newcommand{\TEFF}{\mbox{$T_{\rm eff}$}}
\newcommand{\RSTAR}{\mbox{$R_{\star}$}}
\newcommand{\LOGG}{\mbox{$\log \varg$}}
\newcommand{\MSTAR}{\mbox{$M_{\star}$}}

\newcommand{\VMICRO}{\mbox{$\varv_{\rm micro}$}}

\newcommand{\MSOL}{\mbox{$M_{\sun}$}}
\newcommand{\MSOLPERYR}{\mbox{$M_{\sun}$~yr$^{-1}$}}
\newcommand{\micron}{\mbox{$\mu$m}}
\newcommand{\KMS}{\mbox{km s$^{-1}$}}
\newcommand{\HOH}{\mbox{H$_2$O}}
\newcommand{\PERSQCM}{\mbox{cm$^{-2}$}}
\newcommand{\PERCBCM}{\mbox{cm$^{-3}$}}
\newcommand{\alfcenA}{\mbox{$\alpha$~Cen~A}}
\usepackage{color}
\usepackage{amstext}

\begin{document}

\title{
Spatially resolving the thermally inhomogeneous outer atmosphere of the 
red giant Arcturus in the 2.3 \micron\ CO lines
\thanks{
Based on AMBER observations made with the Very Large Telescope 
and Very Large Telescope Interferometer of the European Southern Observatory. 
Program ID: 092.D-0461(A). 
}
}

\author{K.~Ohnaka\inst{1} 
\and
C.~A.~L.~Morales Mar\'in\inst{1} 
}

\offprints{K.~Ohnaka}

\institute{
Instituto de Astronom\'{i}a, Universidad Cat\'{o}lica del Norte, 
Avenida Angamos 0610, Antofagasta, Chile\\
\email{k1.ohnaka@gmail.com}
}

\date{Received / Accepted }

\abstract
{}
{
The outer atmosphere of K giants shows thermally inhomogeneous structures 
consisting of the hot chromospheric gas and the cool molecular gas. 
We present spectro-interferometric observations of the multicomponent 
outer atmosphere of the well-studied K1.5 giant Arcturus ($\alpha$~Boo) 
in the CO first overtone lines near 2.3~\micron.
}
{
We observed Arcturus with the AMBER instrument at the Very Large Telescope 
Interferometer (VLTI) at 2.28--2.31~\micron\ with a 
spectral resolution of 12\,000 and at projected baselines of 7.3, 14.6, and 
21.8~m. 
}
{
The high spectral resolution of the VLTI/AMBER instrument allowed us to 
spatially resolve Arcturus in the individual CO lines. 
Comparison of the observed interferometric data with the MARCS 
photospheric model shows that the star appears to be significantly larger 
than predicted by the model. It indicates the presence of an extended 
component that is not accounted for by the current photospheric models for this 
well-studied star. 
We found out that the observed AMBER data can be explained by a model 
with two additional CO layers above the photosphere. The inner CO layer is located 
just above the photosphere, at $1.04\pm 0.02$~\RSTAR, with a
temperature of $1600\pm 400$~K and a CO column density of 
$10^{20\pm0.3}$~\PERSQCM. 
On the other hand, 
the outer CO layer is found to be as extended as to $2.6 \pm 0.2$~\RSTAR\ 
with a temperature of $1800 \pm 100$~K and a CO column density of 
$10^{19\pm0.15}$~\PERSQCM. 
}
{
The properties of the inner CO layer are in broad agreement with 
those previously inferred from the spatially unresolved spectroscopic 
analyses. However, our AMBER observations have revealed that the quasi-static 
cool molecular component extends out to 2--3~\RSTAR, 
within which region the chromospheric wind steeply accelerates. 
}


\titlerunning{Spatially resolving the outer atmosphere of Arcturus in the 
2.3~\micron\ CO lines with VLTI/AMBER
}
\authorrunning{Ohnaka \& Morales Mar\'in}
\maketitle

\begin{table*}
\caption {
Summary of the VLTI/AMBER observations of Arcturus and the calibrator \alfcenA. 
}
\begin{center}

\begin{tabular}{l c c c c r l }\hline
\# & $t_{\rm obs}$ & $B_{\rm p}$ & PA     & Seeing   & $\tau_0$ &
${\rm DIT}\times{\rm N}_{\rm f}\times{\rm N}_{\rm exp}$ \\ 
   & (UTC)       & (m)       & (\degr) & (\arcsec)       &  (ms)    &  (ms)  \\
   &             &B2-C1/C1-D0/B2-D0 & B2-C1/C1-D0/B2-D0 & &         &        \\
\hline
\multicolumn{7}{c}{Arcturus: 2014 February 11 (UTC)}\\
\hline
1 & 07:45:14 & 7.3/14.6/21.8 & 21/21/21    & 0.83 & 5.4 & $120\times500\times5$ \\
\hline
\multicolumn{7}{c}{\alfcenA: 2014 February 11 (UTC)}\\
\hline
C1 & 05:02:03 & 10.07/20.17/30.23 & 165/165/165 & 1.02& 4.8 &$120\times500\times5$\\
C2 & 05:36:38 & 10.14/20.32/30.46 & 171/171/171 & 0.97& 5.0&$120\times500\times5$\\
C3 & 06:51:22 & 10.19/20.40/30.59 & 3/3/3 & 0.94& 4.9 &$120\times500\times5$\\
C4 & 07:26:01 & 10.15/20.33/30.48 & 9/9/9 & 0.82& 5.5 &$120\times500\times5$\\
C5 & 08:02:11 & 10.07/20.18/30.25 & 14/14/14 & 1.26& 3.5&$120\times500\times5$\\
\hline
\label{obs_log}
\vspace*{-7mm}

\end{tabular}
\end{center}
\tablefoot{
$B_{\rm p}$: Projected baseline length.  PA: Position angle of the baseline 
vector projected onto the sky. 
DIT: Detector integration time.  $N_{\rm f}$: Number of frames in each 
exposure.  $N_{\rm exp}$: Number of exposures. 
The seeing and the coherence time ($\tau_0$) were measured in the visible. 
}
\end{table*}

\section{Introduction}
\label{sect_intro}

The mass-loss phenomenon is ubiquitous across the Hertzsprung-Russel (H-R) diagram (e.g., Cranmer 
\& Saar \cite{cranmer11}). Despite its importance in stellar evolution and in the chemical enrichment of galaxies, the mass-loss mechanism is not yet 
fully understood in general. 
When Sun-like stars evolve to the red giant branch (RGB) after cessation of 
the hydrogen core burning, the mass-loss rate increases by four 
orders of magnitude or more, from the 
$(2-3) \times 10^{-14}$~\MSOLPERYR\ 
observed in the Sun (Wang \cite{wang98}) to $\ga 10^{-10}$~\MSOLPERYR\ 
(Cranmer \& Saar \cite{cranmer11} and references therein). 

Arcturus ($\alpha$ Boo) is a moderately metal-poor ([Fe/H] = $-0.4$) 
red giant star with 
an initial mass of approximately 1~\MSOL\ (Smith et al. \cite{smith13}) 
and a mass-loss rate 
of $2.5\times 10^{-10}$~\MSOLPERYR\ (Schr\"oder \& Cuntz \cite{schroeder07}). 
The absence of a signature of dust in its infrared spectrum 
suggests that the mass loss is not driven by the radiation pressure on 
dust grains, which is one of the mass-loss mechanisms proposed for 
more evolved stars such as Mira variables in the asymptotic giant branch 
(e.g., H\"ofner \& \mbox{Olofsson} \cite{hoefner18}). 
Moreover, given the very small variability amplitude of Arcturus ($\sim$0.04~mag 
in the visible, Bedding \cite{bedding00}), stellar pulsation is unlikely 
to play a major role in driving the mass loss. 
The Alfv\'en-wave-driven wind is a viable candidate (Suzuki 
\cite{suzuki07}; Airapetian et al. \cite{airapetian10}), given 
the detection of a magnetic field with a surface-averaged strength of 
$\sim$0.5~G (Sennhauser \& Berdyugina \cite{sennhauser11}). 
However, a self-consistent model that 
simultaneously computes the thermal structure and dynamics for the 
weakly ionized atmosphere of red giant stars is lacking so far.

Because of its proximity ($11.3 \pm 0.1$~pc based on a parallax of 
$88.83 \pm 0.54$~mas, van Leeuwen \cite{vanleeuwen07}) and its brightness, 
Arcturus has been well studied from the X-ray to the radio with various 
observational techniques. These observations reveal the complex nature 
of the outer atmosphere, where the stellar wind acceleration is considered 
to take place. On the one hand, the emission lines of ionized metals such 
as Mg II and Ca II indicate a chromosphere with a temperature of close to 
$\sim \! 10^4$~K (e.g., Ayres \& Linskly \cite{ayres75}). 
On the other hand, the observed spectra of the CO fundamental 
lines near 4.7~\micron\ cannot be explained by the chromospheric models 
but suggest the presence of cool molecular gas with temperatures of 
2000--3000~K (Heasley et al. \cite{heasley78}; Wiedemann et al. 
\cite{wiedemann94}), which has been called COmosphere. 
This means that neither the chromosphere nor the cool 
molecular gas covers the entire star, but they presumably coexist in 
spatially inhomogeneous structures. 

The detailed analysis of the infrared spectra of cool evolved stars shows that 
the extended cool molecular outer atmosphere is a characteristics common 
in normal (i.e., non-Mira-type) 
red giants and Mira stars (Hinkle et al. \cite{hinkle78}; 
Tsuji \cite{tsuji88}, \cite{tsuji01}; Tsuji et al. \cite{tsuji97}) 
as well as in red supergiants (RSG, Tsuji \cite{tsuji00a}, \cite{tsuji00b}). 
Tsuji (\cite{tsuji00b}) coined  the word ``MOLsphere'' for this cool extended 
component, which extends out to a few \RSTAR\ with temperatures of 
1000--2000~K (see Tsuji \cite{tsuji09} for a discussion of whether the COmosphere 
and MOLsphere are identical).

Infrared interferometry has enabled us to spatially resolve 
this cool molecular component of the outer atmosphere. 
On the one hand, the interferometric observations of the outer atmosphere of 
Mira stars can be reasonably explained by dynamical model atmospheres 
(Woodruff et al. \cite{woodruff09}; Hillen et al. \cite{hillen12}; 
Wittkowski et al. \cite{wittkowski08}, \cite{wittkowski11}, 
\cite{wittkowski16}). 
On the other hand, the spatially resolved 
observations of normal red giants reveal that the stars appear to be 
significantly more extended than predicted by current hydrostatic 
photospheric models, which are considered to be adequate for these stars 
with small variability amplitudes (Mart\'i-Vidal et al. \cite{marti-vidal11}; 
Ohnaka et al. \cite{ohnaka12}; Ohnaka \cite{ohnaka13b}). 

The spectro-interferometric observations of the 2.3 \micron\ CO lines 
in the K5 giant Aldebaran ($\alpha$ Tau) revealed that the molecular 
atmosphere extends out to $\sim$2.5 \RSTAR\ even in a star as 
warm as 3900~K (Ohnaka \cite{ohnaka13b}). 
In this paper, we present spectro-interferometric 
observations of the CO lines in the K1.5 giant Arcturus ($\alpha$ Boo), 
which is even warmer than Aldebaran. 
We describe the observations and data reduction in 
Sect.~\ref{sect_obs} and the observational results in Sect.~\ref{sect_res}. 
The modeling of the data is presented in Sect.~\ref{sect_model}, followed 
by a discussion (Sect.~\ref{sect_discuss}) and concluding remarks 
(Sect.~\ref{sect_concl}).

\begin{figure*}
\resizebox{\hsize}{!}{\rotatebox{0}{\includegraphics{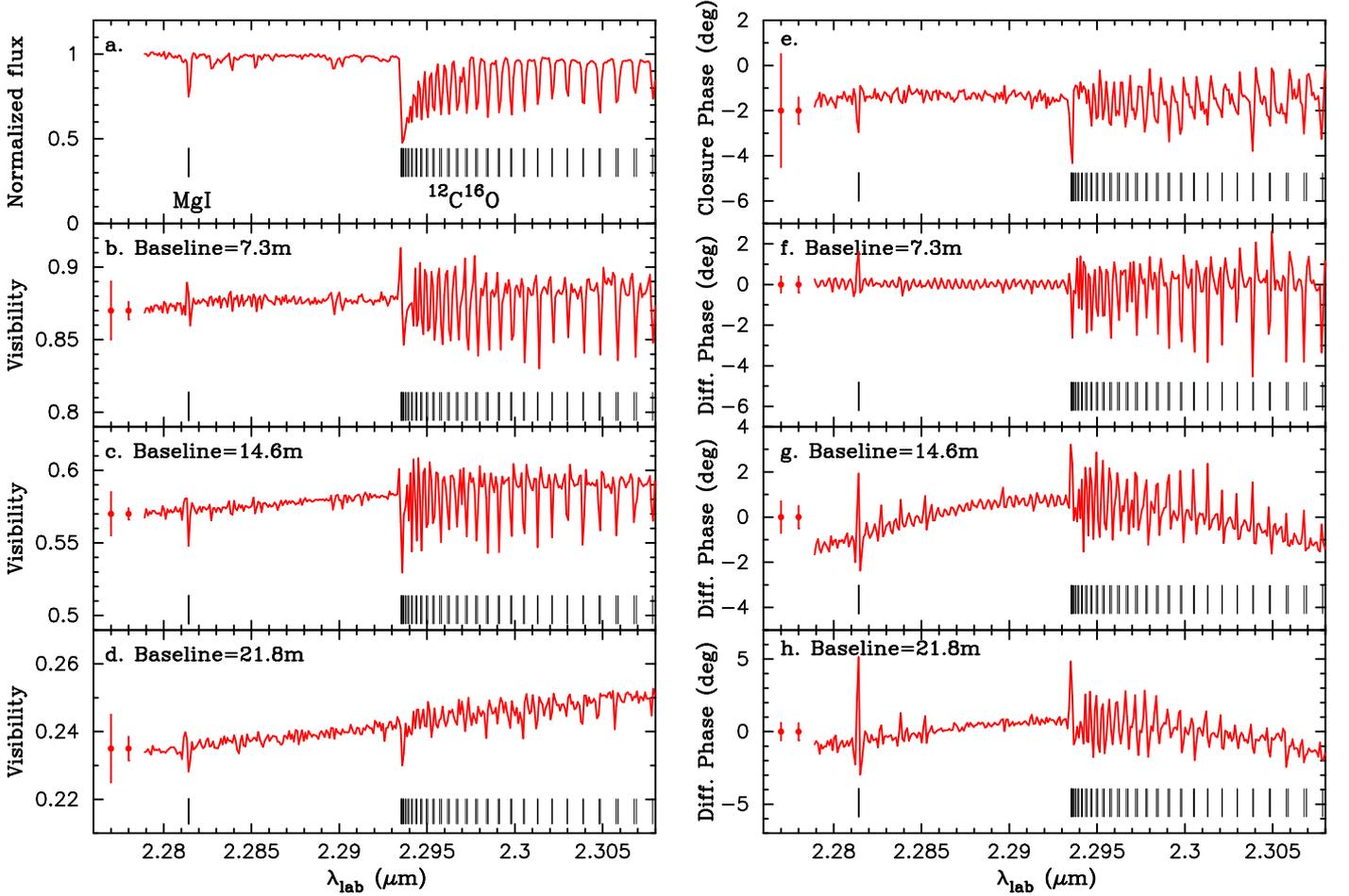}}}
\caption{
VLTI/AMBER observations of the red giant Arcturus. 
{\bf a:} Observed spectrum normalized to the continuum. The positions 
of the CO lines and the Mg I line are marked with the ticks. 
{\bf b--d:} Visibilities observed at 7.3, 14.6, and 21.8~m, 
respectively. The leftmost error bars represent the typical total errors, 
while the smaller error bars represent the typical errors without the systematic 
errors resulting from the absolute calibration. 
{\bf e:} Observed closure phase. The errors are shown in the same manner 
as in panels {\bf b}--{\bf d}. 
{\bf f--h:} Differential phases observed at the 7.3, 14.6, and 21.8~m 
baselines, respectively. The errors are shown in the same manner 
as in panels {\bf b}--{\bf d}. 
}
\label{obsres}
\end{figure*}

\section{VLTI/AMBER observations}
\label{sect_obs}

We observed Arcturus with the near-infrared interferometric instrument 
AMBER (Petrov et al. \cite{petrov07}) at ESO's Very Large Telescope 
Interferometer (VLTI). The AMBER instrument operates between 1.3 and 
2.4~\micron\ and allows us to combine three 8.2~m unit telescopes (UTs) or 
1.8~m auxiliary telescopes (ATs) and achieve a spatial resolution of 
down to 2~mas at 2~\micron\ with spectral resolutions of 35, 1500, and 
12\,000. 
Our AMBER observations of Arcturus took place on 2014 February 11 (UTC) 
using the B2-C1-D0 AT configuration, which provided projected baseline
lengths of 7.3, 14.6, and 21.8~m. 
We observed 
a spectral window between 2.28 and 2.31~\micron\ with a spectral resolution 
of 12\,000, which is sufficient to spectrally resolve individual CO 
first overtone lines near the $\varv = 2 - 0$ band head located at 
2.294~\micron. 
We did not use the VLTI fringe-tracker FINITO, because Arcturus saturates 
FINITO. Nevertheless, the high brightness of the target enabled us to 
record fringes of good quality with a detector integration time (DIT) 
of 120~ms. The log of our observations is summarized in Table~\ref{obs_log}.

The data were reduced with amdlib 
ver~3.0.7\footnote{http://www.jmmc.fr/data\_processing\_amber.htm}. The amdlib 
software extracts the following observables from the recorded interferograms 
based on the P2VM algorithm (Tatulli et al. \cite{tatulli07}; Chelli et al. 
\cite{chelli09}): squared visibility amplitude, closure phase (CP), 
and wavelength-differential phase (DP). The visibility amplitude 
corresponds to 
the amplitude of the complex Fourier transform of the object's intensity 
distribution on the sky. 
The CP is the sum of the phases of the complex Fourier transform 
measured on three baselines forming a closed telescope triangle.  
Since the CP is zero or $\pi$ for a point-symmetric object, 
non-zero and non-$\pi$ CPs indicate asymmetry in the object. 
The DP represents the Fourier phase of the object on a given baseline 
in spectral features (e.g., atomic or molecular lines) with respect to 
the continuum. The DP is related to the photocenter shift of the object 
in spectral features with respect to the continuum.

The interferometric observables extracted with amdlib were calibrated 
with the observations of \alfcenA\ (G2V), which was observed 
not only for the calibration of the Arcturus data, but 
also for other targets in the same observing program. 
We computed the transfer function adopting 
a uniform-disk diameter of $8.314 \pm 0.016$~mas for \alfcenA\ 
(Kervella et al. \cite{kervella03}). 
The Arcturus data were calibrated with the transfer function values measured 
before and after Arcturus, while the errors in the interferometric 
observables were estimated including the standard deviation of all 
transfer function values measured throughout the night. 
The atmospheric conditions were very stable with a seeing of mostly 
0\farcs8--1\farcs0 and coherence time ($\tau_0$) of 4--5~ms, 
and therefore the transfer function was stable throughout the night, as shown in Fig.~\ref{tfplot}. 
This means that although the DIT is longer than the coherence time 
(20--30~ms at 2.3~\micron, assuming $\tau_0(\lambda) \propto \lambda^{6/5}$), 
the effects of the atmosphere on the interferometric observables were 
stable, and therefore the science data can be well calibrated.

We compared the results we obtained by keeping only the best 20\% and 80\% of the 
frames in terms of the fringe signal-to-noise ratio (S/N). The results obtained with the best 20\% 
of the frames show smaller errors in the visibilities. On the other hand, 
the errors in the DPs and CP are smaller if the best 80\% of the frames are 
kept (i.e., discarding the worst 20\% of the frames). Therefore, we adopted 
the visibilities obtained from the best 20\% of the frames and the CP and DPs 
obtained from the best 80\% of the frames. The spectrum was obtained by 
using the best 80\% of the frames. 

The total errors in the visibilities, DPs, and CP are dominated by the 
systematic errors resulting from the absolute calibration. 
We estimated the relative errors in the wavelength-differential 
visibilities, DPs, and CP (i.e., errors without the systematic errors 
due to the absolute calibration) as follows. 
The data of Arcturus consist of five exposures 
taken consecutively (see Table~\ref{obs_log}). We reduced and calibrated 
each exposure separately in the same manner as described above. 
The visibilities resulting from five exposures are normalized to 1 
in the continuum. We took the standard deviation of these normalized 
visibilities among 
five exposures at each wavelength as the relative errors in the 
wavelength-differential visibilities. The relative errors in the 
wavelength-differential DPs and CP were estimated in the same manner, but 
without normalizing to 1.

The wavelength calibration was carried out using the telluric lines 
identified in the observed spectrum of \alfcenA, 
as described in Ohnaka et al. (\cite{ohnaka09}). The uncertainty in 
wavelength calibration is $2.1\times10^{-5}$~\micron, 
which translates into 2.7~\KMS. 
The observed wavelength scale was then converted into the laboratory frame 
using the heliocentric velocity of $-5.2$~\KMS\ of Arcturus 
(Gontscharov \cite{gontscharov06}).
The spectroscopic calibration (e.g., removal of the telluric lines and 
instrumental effects from the observed spectrum of Arcturus) was done 
using the observed spectrum of \alfcenA\ based on 
the procedure described in Ohnaka et al. (\cite{ohnaka13a}).

\section{Results}
\label{sect_res}

Figure~\ref{obsres} shows the observed spectrum, visibilities, closure phase, 
and differential phases of Arcturus\footnote{Strictly speaking, taking 
the square root of the squared visibility amplitude extracted with 
amdlib makes the errors asymmetric. However, this effect is very small 
in our case, and the errors in the visibility amplitude are still nearly 
symmetric. We show the visibility amplitude to facilitate a comparison with 
our previous results.}. 
The high spectral resolution 
of the AMBER instrument enables the signatures of the individual 
CO lines\footnote{These
  spectral features that appear to be single lines consist of two CO 
transitions with high $J$ and low $J$, where $J$ is the rotational quantum 
number.} to be 
clearly visible in the interferometric observables. 
It should be noted that while the changes in the visibilities observed 
in the CO lines may appear to be marginal compared to the error bars 
(the larger error bars shown in the far left in 
Figs.~\ref{obsres}b--\ref{obsres}d), 
the errors are dominated by the systematic errors in the absolute 
calibration. 
The relative errors in the wavelength-differential visibilities 
are much smaller, 0.7\% at the shortest and middle baselines and 
1.5\% at the longest baseline, 
as shown by the smaller error bars in the figure. 
This means that the signatures of the CO lines in the visibilities are real. 

\begin{figure}
\resizebox{\hsize}{!}{\rotatebox{0}{\includegraphics{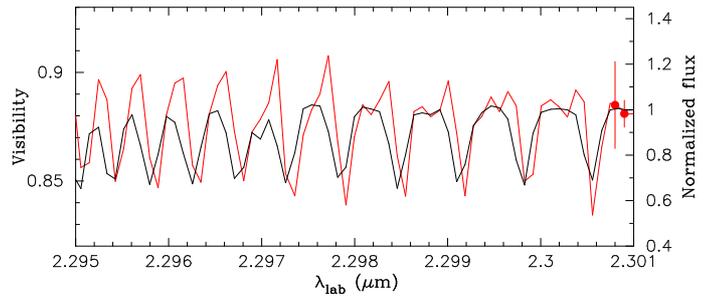}}}
\caption{
Enlarged view of the visibility in the CO lines 
observed at the shortest baseline of 7.3~m (red line). 
It shows asymmetry with respect to the line center. 
The typical errors are shown in the right in the same manner as in 
Fig.~\ref{obsres}. 
The scaled observed spectrum is shown with the black line. 
}
\label{obsresCO_vis_alfboo}
\end{figure}

The drop in visibilities in the CO lines means that 
the star appears larger in the CO lines. 
This is qualitatively expected because the CO first overtone lines form 
in the upper photosphere and in the outer atmosphere. 
However, the visibilities in the CO lines near the band head 
($\la$2.3~\micron) are higher 
than in the continuum. This is particularly clearly visible in the data 
taken at the shortest baseline, as shown in Fig.~\ref{obsresCO_vis_alfboo}. 
The figure also reveals that the visibilities in these CO lines do not 
simply rise above the continuum level, but are asymmetric with respect 
to the center of the line profile: the visibility shows a spike above the 
continuum level in the blue wing, while it drops below the continuum level 
in the red wing. As we show in Sect.~\ref{sect_model}, the visibility 
spikes rising above the continuum level cannot be reproduced by our models. 
The visibilities higher than the continuum level mean that the star appears 
smaller in the blue wing of the lines. A possible reason is a strong 
stellar spot. 
Similar visibilities that are asymmetric in the blue and red wing (i.e., 
$\text{tilde}$-shaped) are observed in the RSGs Betelgeuse and Antares 
(Ohnaka et al. \cite{ohnaka09}, \cite{ohnaka11}, \cite{ohnaka13a}), 
and can be explained by upwelling or downdrafting large spots in the 
atmosphere. 
The inhomogeneous velocity field in the atmosphere of Antares 
has been imaged with VLTI/AMBER (Ohnaka et al. \cite{ohnaka17}).  
Therefore, Arcturus might also show such spots. Further observations, 
preferably aperture-synthesis imaging, is necessary to confirm this.

\begin{figure}
\resizebox{\hsize}{!}{\rotatebox{0}{\includegraphics{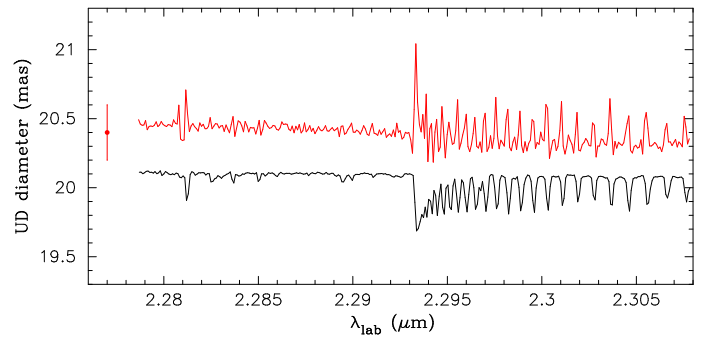}}}
\caption{
Uniform-disk diameter of Arcturus. The typical error is shown in the 
left. The scaled observed spectrum is shown in black. 
}
\label{udfit}
\end{figure}

We fit the visibilities measured at each wavelength with a uniform disk 
to have an approximate idea about the increase of the star's size in the 
CO lines. As Fig.~\ref{udfit} shows, the uniform-disk diameter in the 
continuum is $20.4 \pm 0.2$~mas, 
which agrees well with the previous measurements in the infrared 
($20.44 \pm 0.16$~mas at 2.22~\micron, Verhoelst et al. \cite{verhoelst05}; 
$20.304 \pm 0.011$~mas at 1.65~\micron, Lacour et al. \cite{lacour08}).  
The uniform-disk diameter increases up to $\sim$21~mas in the CO band 
head, and it stays at $\sim$20.6~mas in the individual CO lines. 
Although the increase in the uniform-disk diameter in the CO band head and 
CO lines may seem insignificant, it provides us with important information 
about the extended atmosphere, as we present in Sect.~\ref{sect_model}. 
In the CO lines near the band head, the uniform-disk diameter decreases to 
$\sim$20.2~mas, reflecting the visibility spikes described above.

The continuum visibilities observed at the 14.6 and 21.8~m baselines 
show a slight increase with wavelength.  This is for the following reason. 
On the one hand, the spatial frequency decreases toward longer wavelengths 
($\lambda$) for a given baseline length ($B$), because it is defined as 
$B/\lambda$. On the other hand, if the star approximately appears to be 
a uniform disk or limb-darkened disk in the continuum, the visibility 
increases toward lower spatial frequencies in the first lobe 
(as in the case of our observations). 
Therefore, the visibility in the continuum, 
when shown as a function of wavelength, increases toward longer wavelengths. 

\begin{figure}
\resizebox{\hsize}{!}{\rotatebox{0}{\includegraphics{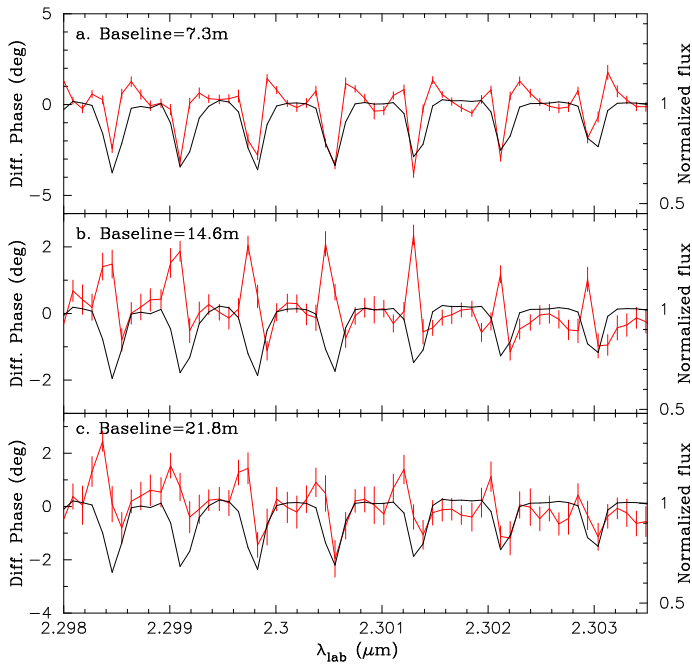}}}
\caption{
Enlarged view of the observed differential phases across the CO line 
profiles. The data obtained at 7.3, 14.6, and 21.8~m are shown 
in panels {\bf a}, {\bf b}, and {\bf c}, respectively. 
In each panel, the red line with the error bars (total errors) 
shows the observed 
differential phase, while the black line shows the observed scaled spectrum. 
}
\label{obsresCO}
\end{figure}

We also detected non-zero DPs and CP in the CO lines, which indicate 
asymmetry in the CO lines, 
perhaps due to inhomogeneities in the CO-line-forming 
upper layers, as mentioned above. 
Furthermore, 
a closer look at the data shows that the DPs observed at the 14.6 and 21.8~m 
baselines are asymmetric with respect to the center of the line profile. 
As Fig.~\ref{obsresCO} shows, the maxima of the DPs are located in 
the blue wing of the line profiles, while the minima are located in the 
red wing. 
This means that the photocenter of the object is different in the blue and 
red wings. In other words, the star appears differently in the blue and red 
wings. The DP observed on the shortest baseline of 7.3~m also shows 
signatures of such asymmetry across the line profile (the minima of the 
visibilities are slightly shifted to the blue wing), but less significant 
compared to two longer baselines. 
This asymmetry in the DPs and CPs across the CO line 
profiles is similar to what has been observed in the RSGs 
Betelgeuse and Antares (Ohnaka et al. \cite{ohnaka09}, \cite{ohnaka11}, 
\cite{ohnaka13a}) and may also imply inhomogeneous structures, together 
with the asymmetry in the visibility observed at the 7.3~m 
baseline, as described above.

The weak line at 2.2814~\micron\ is identified to be Mg I. 
The signature of this line can be seen in the observed visibilities as well as 
in the CP and DPs. 
The visibilities in the Mg I line at the shortest and longest 
baselines show a $\text{tilde}$-shaped signature: a peak in the blue wing, and a 
drop in the red wing of the line. The uniform-disk diameter in the Mg I line 
also reflects the asymmetry, as shown in Fig.~\ref{udfit}. 
This is similar to the visibility spikes 
in the CO lines described above, and cannot be 
interpreted by the modeling we present here. 
Therefore, we refrain from modeling the Mg I line in this work.

There are also other very weak lines shortward of the CO band head. However, 
they are located at the wavelengths of some telluric lines, and 
it is possible that they are still affected by the residual of the removal 
of the telluric lines.

\begin{figure}
\resizebox{\hsize}{!}{\rotatebox{0}{\includegraphics{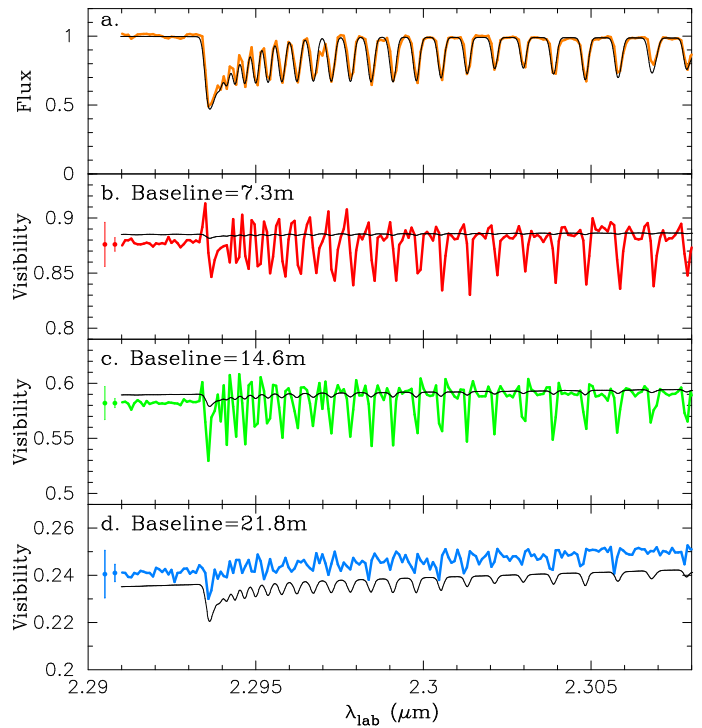}}}
\caption{
Comparison of the observed visibilities of Arcturus with those predicted 
by the MARCS photospheric model. Panel {\bf a} shows a comparison 
of the spectrum, while panels {\bf b}--{\bf d} show a comparison 
of the visibilities observed at the 7.3, 14.6, and 21.8~m baselines, 
respectively. 
In each panel, the thick solid line represents the observed data, and 
the thin solid line represents the model. 
The typical errors are shown in the same manner as in Fig.~\ref{obsres}. 
The parameters of the MARCS 
model are described in Sect.~\ref{subsect_marcs}. 
}
\label{model_marcs}
\end{figure}

\section{Modeling the AMBER data}
\label{sect_model}

\subsection{Comparison with MARCS models}
\label{subsect_marcs}

We first compare the observed visibilities with the MARCS photospheric models 
(Gustafsson \cite{gustafsson08}) to see whether the observed data 
in the CO lines can be explained by the 
photosphere without any additional component such as the MOLsphere or COmosphere. 
The MARCS models are plane-parallel or spherical hydrostatic photospheric 
models that incorporate a great number of atomic and molecular lines. 
The models are based on local thermodynamical equilibrium (LTE), in which  
convection is taken into account by means of the mixing length theory. 
In the spherical MARCS models, 
each model is specified by effective temperature (\TEFF), surface gravity 
($\log \varg$), microturbulent velocity (\VMICRO), stellar mass (\MSTAR), 
and chemical composition.

For the basic stellar parameters of Arcturus, 
we adopted \TEFF\ = $4250\pm 50$~K, \LOGG\ = $1.7 \pm 0.1$, \MSTAR\ = 
$1.1$~\MSOL, \VMICRO\ = 2.0~\KMS, and [Fe/H] = $-0.4$ (Tsuji \cite{tsuji09}; 
Smith et al. \cite{smith13}). 
The CNO abundances in Arcturus roughly agree with the moderately CN-cycled 
chemical composition used in the MARCS model grid. Therefore, we used 
the MARCS model with \TEFF\ = 4250~K, \LOGG\ = 1.5, \MSTAR\ = 1.0~\MSOL, 
\VMICRO\ = 2.0~\KMS, and [Fe/H] = $-0.5$ with the moderately CN-cycled 
composition. Using the density and temperature stratifications downloaded 
from the MARCS model site\footnote{http://marcs.astro.uu.se}, we first 
computed the intensity profile at each wavelength and then the flux and 
visibility as described in Ohnaka (\cite{ohnaka13b}). 
The angular scale of the model visibilities was set so that the model 
visibilities computed in the continuum at three observed baselines matched the 
observed data within the measurement errors. 
The CO line opacity 
was calculated using the line list of Goorvitch (\cite{goorvitch94}), 
and a Gaussian line profile was assumed. 
The flux and visibility were spectrally convolved to match the 
 spectral resolution of AMBER, 12\,000. 

Figure~\ref{model_marcs} shows a comparison of the MARCS model and 
the observed data of Arcturus. 
While there is an offset between the observed and model visibility levels 
at the longest baseline of 21.8~m, this is within the uncertainty of the absolute 
visibility calibration. 
The model can clearly reproduce the observed spectrum very well. 
However, the visibilities predicted by the model show too shallow 
decreases in the CO lines except for the 21.8~m baseline. 
This means that the atmosphere of Arcturus is 
more extended than the MARCS model predicts, or there is an extended component 
above the photosphere that is not accounted for by the MARCS model. 

We also computed the synthetic spectrum and visibilities using the MARCS 
model with a higher turbulent velocity of 5~\KMS\ instead of 2~\KMS\ 
(the other model parameters were the same). The model with 5~\KMS\ predicts the 
CO absorption lines to be slightly stronger (therefore, the agreement with 
the observed spectrum is slightly worse). The predicted visibilities are 
nearly the same as those predicted by the model with 2~\KMS : the model 
visibilities in the CO lines are still too high compared to the observed 
data. 
The geometrical extension of these MARCS photospheric models is only 
$\sim$2\%, which is insufficient by far
to explain the observed visibilities. 
Our result 
is qualitatively consistent with the finding of Tsuji (\cite{tsuji09}) 
that the strong CO first overtone lines in Arcturus cannot be explained 
by current photospheric models and show the presence of the additional 
extended component MOLsphere. 
In other words, our AMBER observations have spatially resolved the MOLsphere 
of Arcturus in the individual CO lines.

\begin{figure}
\resizebox{\hsize}{!}{\rotatebox{0}{\includegraphics{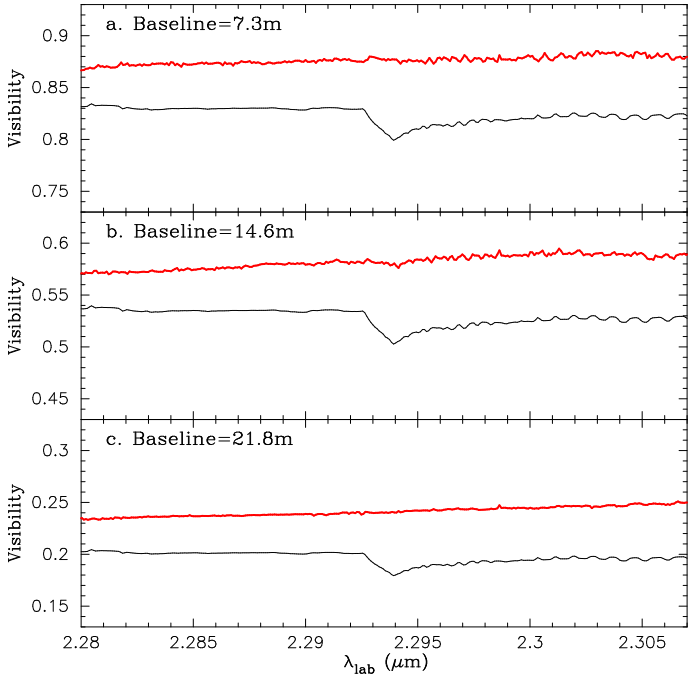}}}
\caption{
Visibilities of Arcturus obtained at a spectral resolution of 1600. 
The visibilities at the 7.3, 14.6, and 21.8~m baselines are plotted with thick red lines in panels {\bf a}, {\bf b}, and {\bf c}, respectively. 
In each panel, the black line shows the scaled spectrum. 
}
\label{obsresR1600}
\end{figure}

The result that the MARCS model cannot explain the visibilities 
observed in the individual CO first overtone lines of normal 
(i.e., non-Mira-type) K--M giants has been reported for the K5 giant 
Aldebaran and M7 giant BK~Vir (Ohnaka \cite{ohnaka13b}; Ohnaka et al. 
\cite{ohnaka12}), which are cooler than Arcturus. 
However, Arroyo-Torres et al. (\cite{arroyo-torres14}) found no signatures 
of a non-photospheric component in the visibilities obtained across 
the CO bands at 2.3--2.45~\micron\ 
in four out of five red giants from G8III to M6III. 
Only in the M2III star $\beta$~Peg did they find signatures of a
non-photospheric extended component. 
However, their negative detection may be due to the lower spectral resolution 
of 1500 compared to the 12\,000 used in our observations. 
To see whether the signatures of the CO lines in the visibilities observed 
in Arcturus can still be detected with the spectral resolution of 1500, 
we spectrally binned the raw data (object, dark, sky, and P2VM calibration 
data) to a resolution of 1600 with a running box-car function as described 
in Ohnaka et al. (\cite{ohnaka09}). The binned data were reduced and 
calibrated in the same manner as the original unbinned data. 
Figure~\ref{obsresR1600} illustrates that while the spectrum still shows 
the CO band head, the visibilities show no signatures of the CO band when 
binned down to 1600. 
Therefore, high spectral resolution is essential to confirm the presence 
or absence of the MOLsphere in normal K--M giants. 

In contrast to the MOLsphere of normal (non-Mira-type) K--M giants, 
the AMBER observations of 
a sample of RSGs carried out by Arroyo-Torres et al. 
(\cite{arroyo-torres15}) and Wittkowski et al. (\cite{wittkowski17}) 
show the presence of the MOLsphere, although the same spectral resolution of 
1500 was used. 
Arroyo-Torres et al. (\cite{arroyo-torres15}) showed that the visibilities 
in the CO bands observed in their RSGs significantly 
deviate from what is predicted by the photospheric models, which is the 
signature of the presence of the MOLsphere. On the other hand, most of 
the red giants in their sample do not show the signature of the MOLsphere 
(see their Fig.~10). 
Therefore, it is suggested that the emission from the MOLsphere in RSGs is 
stronger than in normal K--M giants, possibly because the 
density and/or temperature of the MOLsphere is higher and/or the MOLsphere 
is more extended.

\begin{table*}
\begin{center}
\caption {Parameters of the MARCS+MOLsphere models for Arcturus and the best-fit 
solution}
\vspace*{-2mm}

\begin{tabular}{l l l}\hline
Parameter &  Searched range &  Solution \\
\hline
\rule{0pt}{1.0\normalbaselineskip}CO column density of the inner layer: $N_{\rm inner}$ (\PERSQCM) & 
$10^{16}, 5\times10^{16}, 10^{17}, ..., 10^{22}$ 
& $10^{20\pm0 .3}$ \\
CO column density of the outer layer: $N_{\rm outer}$ (\PERSQCM) & 
$10^{16}, 5\times10^{16}, 10^{16}, ..., 10^{22}$ 
& $10^{19\pm 0.15}$ \\
Temperature of the inner layer: $T_{\rm inner}$ (K) &
1000 ... 2500 ($\Delta T_{\rm inner}$ = 100~K) & $1600 \pm 400$\\
Temperature of the outer layer: $T_{\rm outer}$ (K) &
1000 ... 2500 ($\Delta T_{\rm outer}$ = 100~K) & $1800 \pm 100$\\
Inner radius the inner layer: $R_{\rm inner}$ (\RSTAR) &
1.02, 1.03, 1.04, ..., 1.1 (thickness = 0.01 ... 0.02~\RSTAR)& $1.04 \pm 0.02$\\
Inner radius the outer layer: $R_{\rm outer}$ (\RSTAR) &
1.1, 1.5, 2.0, ... , 4.0 (thickness = 0.1~\RSTAR) & $2.6\pm 0.2$\\
\hline
\label{table_param}
\vspace*{-7mm}

\end{tabular}
\end{center}
\end{table*}

\begin{figure}
\resizebox{\hsize}{!}{\rotatebox{0}{\includegraphics{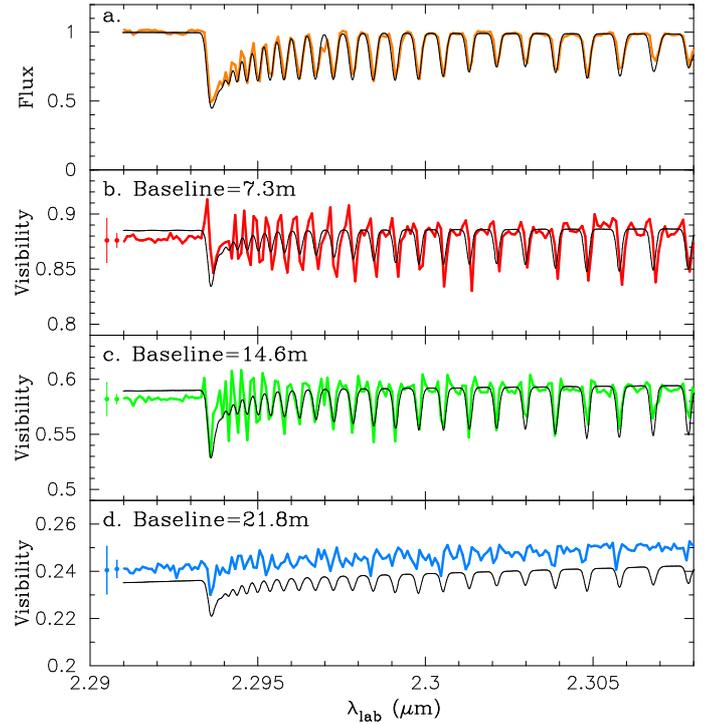}}}
\caption{
Comparison of the observed visibilities of Arcturus with those predicted 
by the MARCS+MOLsphere model, shown in the same manner as 
in Fig.~\ref{model_marcs}.
The model parameters are described in Sect.~\ref{subsect_marcs_wme} and listed 
in Table~\ref{table_param}. 
}
\label{model_marcs_wme}
\end{figure}

\subsection{Comparison with MARCS+MOLsphere models}
\label{subsect_marcs_wme}

As described in Sect~\ref{sect_intro}, the dust-driven stellar wind models and 
pulsation-driven wind models are probably not appropriate for Arcturus. 
No Alfv\'en-wave-driven models currently take the thermal structure and dynamics of the weakly ionized 
outer atmosphere of red giants into account in a self-consistent manner. 
In other words, it is not clear yet which physical process is responsible for 
the formation of the MOLsphere. 

Therefore, 
to characterize the extended MOLsphere component 
detected in our AMBER observations, 
we used the semi-empirical model that has been applied 
to derive the parameters of the MOLsphere in our previous studies 
(Ohnaka et al. \cite{ohnaka12}; Ohnaka \cite{ohnaka13b}). 
In this model, one or two layers are added above the MARCS photospheric 
models. 
Each layer is defined by its inner and outer radius and 
assumed to have a constant temperature and density. 
We adopted the photospheric microturbulent velocity of 2~\KMS\ for these 
layers. We first attempted to explain the observed data with the 
MARCS+1-layer models. However, it turned out to be impossible to obtain a 
satisfactory fit to the visibilities measured on all three baselines. 
Therefore, we carried out the modeling with two layers. 
The ranges of the parameters are listed in Table~\ref{table_param}, 
together with the best-fit solution. 
We note that when computing the model grid, 
the radius of the inner layer was always set to be equal to or larger than 
the uppermost layer of the MARCS model, which is located at 1.02~\RSTAR. 
The uncertainties in the parameters of the best-fit model 
were estimated by varying them around 
the best-fit solution. 

The geometrical thickness 
of the layers was not well constrained in our previous studies, 
and therefore was assumed to be 0.1~\RSTAR. However, in the case of 
Arcturus, the models with this assumption did not reproduce the observed
data even with two layers. 
Therefore, we changed the thickness of the layers as well. 
We found that it is necessary to reduce the thickness of the inner layer 
at least to 0.02~\RSTAR\ to obtain a reasonable fit to the data. 
Models with thicknesses of between 0.01 and 0.02~\RSTAR\ can reproduce the 
data equally well, which means that we cannot further constrain the thickness 
of the inner layer. 
On the other hand, the thickness of the outer layer is not well constrained, 
and the adoption of 0.1~\RSTAR\ resulted in reasonable agreement with the 
data.

Figure~\ref{model_marcs_wme} shows a comparison between the best-fit model and 
the observed data. 
The observed visibilities in the CO lines are much better reproduced than 
with the MARCS-only model presented in Sect.~\ref{subsect_marcs}. 
The agreement with the observed spectrum is also reasonable. 
The visibility decrease observed in the CO lines at the shortest 
baseline of 7.3~m suggests the presence of a very extended component. 
This is why our modeling shows that the outer layer extends out to 
2.6~\RSTAR. 
The inner layer is dense and is located very close to the star, at 
1.04~\RSTAR.  
The presence of this dense compact layer is needed to explain the observed 
data for the following reason. 
If the object only consists of the photosphere and the outer CO layer, 
the visibility predicted in the CO lines at the longest baseline is 
significantly lower than that observed, owing to the extended outer CO layer. 
The emission from the 
dense inner CO layer just above the photosphere causes the object to appear 
more compact, canceling the effect described above due to the outer CO layer. 

We estimated the gas density in the MOLsphere CO layers as follows. 
We can convert the derived CO column density into the CO number density 
by dividing by the geometrical thickness of the layer. 
While we assumed two distinct layers to simplify the models, the actual 
density distribution of the MOLsphere can be continuous. 
Therefore, we took 
$(R_{\rm inner\, (outer)} - \RSTAR )$ as an upper limit on the geometrical 
thickness of the CO layer. 
The CO number densities 
derived for the inner and outer CO layers are $1.2 \times 10^{9}$ and 
$3.1 \times 10^{6}$~\PERCBCM, respectively. These values are 
lower limits of the CO number densities. 
Then assuming chemical equilibrium at the temperature derived for each 
CO layer, we estimated the H$_2$, H, and He number densities so that 
the CO number density is reproduced. The gas density, which can be 
computed from the derived H$_2$, H, and He number densities, is 
$1.5 \times 10^{-11}$ and $3.7 \times 10^{-14}$~g~cm$^{-3}$ for the inner 
and outer CO layer, respectively. 

It may appear to be puzzling that the spectrum predicted by the 
MARCS+MOLsphere model is nearly the same as that from the MARCS-only model. 
This can be explained as follows. The spectrum from the extended MOLsphere 
beyond the limb of the star shows the CO lines in emission. 
However, the spectrum inside the limb shows additional absorption in the 
CO lines, because we see the cooler MOLsphere in front of the warmer 
star. In the total spectrum, which is the spatial sum of these emission 
and absorption spectra, the signatures of the MOLsphere can disappear, 
as illustrated in Ohnaka (\cite{ohnaka13b}, Sect.~4). 

Ryde et al. (\cite{ryde02}) detected \HOH\ lines near 12~\micron\ in 
Arcturus, which is a surprise because the photosphere of Arcturus is 
deemed to be too hot for \HOH\ to form. Ryde et al. (\cite{ryde02}) 
argued that the \HOH\ lines originate from the uppermost photospheric 
layers. They showed that the observed \HOH\ lines can be explained without 
introducing a non-photospheric component such as the MOLsphere 
if the temperature of the uppermost 
photospheric layers is decreased by $\sim$300~K. 
While this scenario for the formation of \HOH\ may still be viable, 
it does not disprove the presence of the MOLsphere either, because it is now 
spatially resolved with our VLTI/AMBER observations.

\begin{figure*}
\sidecaption
\rotatebox{0}{\includegraphics[width=12cm]{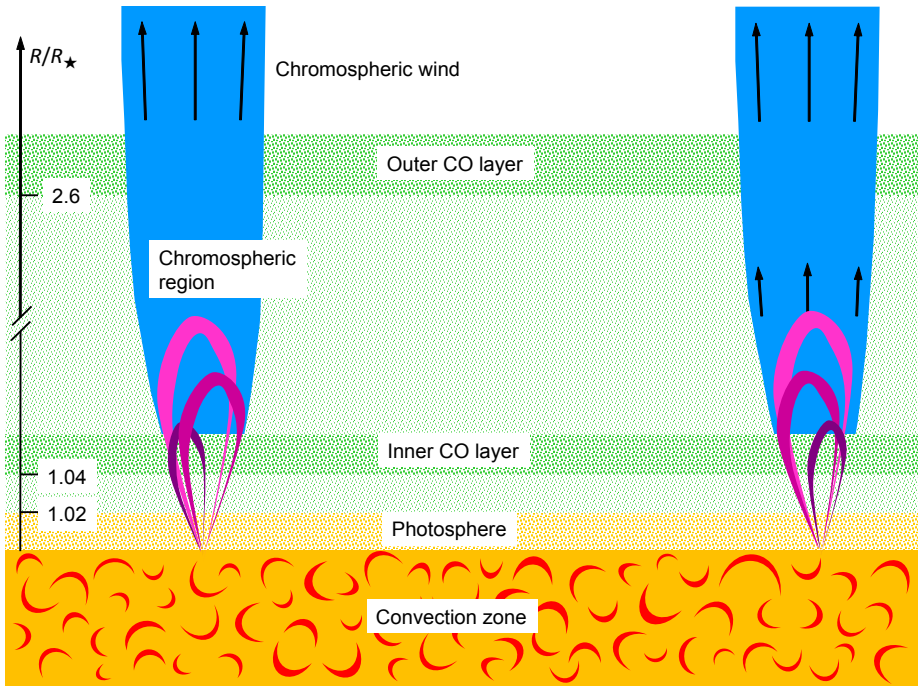}}
\caption{
Schematic view of the thermally and dynamically inhomogeneous outer 
atmosphere of Arcturus. The structure of the chromospheric region is 
based on Fig.~6 of Ayres et al. (\cite{ayres03}). 
The authors postulated that closed magnetic loops (shown in violet), 
which are responsible for the hot coronal gas, 
are submerged in the chromospheric gas at lower temperatures 
(shown in blue). 
The locations of the two CO layers are based on our modeling. 
While we adopted two discrete CO layers in our modeling, it is possible 
that the actual density distribution of this cool molecular component is 
continuous. 
This is represented by the pale green region between the inner and outer 
CO layers as well as between the photosphere and the inner CO layer. 
}
\label{alfboo_schematic}
\end{figure*}

\section{Discussion}
\label{sect_discuss}

The analysis of the spectra of the CO fundamental lines of a sample of 
red giant stars, including Arcturus, 
by Wiedemann et al. (\cite{wiedemann94}) suggests a monotonic 
decrease in temperature above the temperature minimum predicted by the 
chromospheric models. 
The authors estimated temperatures of 2000--3000~K in the same atmospheric heights 
where the chromospheric model predicts the temperature to increase up to 
$\sim$6000~K. 
The temperatures of the cool component 
are in broad agreement with the temperatures of the MOLsphere derived 
from our modeling, given the differences in the observed data as well as 
in the photospheric models. 

Tsuji (\cite{tsuji09}) estimated the temperature and CO column density 
of the MOLsphere of Arcturus to be 2000~K and $5\times10^{19}$~\PERSQCM, 
respectively, based on the spectrum of the strong CO fundamental lines. 
These values agree with the results that we obtained for the inner layer, 
$1600\pm400$~K and $(0.5-2.0)\times10^{20}$~\PERSQCM, given the uncertainties in 
both models. 
Tsuji (\cite{tsuji09}) assumed the inner and outer radius of the 
MOLsphere to be 1.04 and 1.11~\RSTAR, respectively, which also agrees fairly 
well with the inner radius of 1.04~\RSTAR\ 
that we derived for the inner CO layer. The modeling of Tsuji (\cite{tsuji09}) 
does not include a second, more extended layer, because he considered a 
model with as few free parameters as possible to explain the 
spatially unresolved spectra (albeit of much higher spectral resolution)
of the CO fundamental lines.

Based on the tentative detection of X-ray emission from 
Arcturus and the analysis of UV emission lines, Ayres et al. (\cite{ayres03}) 
postulated that the hot coronal gas with temperatures of a few $10^5$~K 
associated with closed magnetic loops is 
submerged (``buried alive'' , as the authors expressed it) in the chromosphere 
with a temperature below $\sim \!\! 10^4$~K. Furthermore, they concluded that 
there should be a cool molecular layer below the chromosphere. 
Figure~6 of Ayres et al. (\cite{ayres03}) shows a cool molecular layer, 
which is presented as COmosphere in their figure, very close to the star, 
within $\sim$1.05~\RSTAR. This agrees with the radius of the inner CO layer of 
our model. They also pointed out the possible presence of inhomogeneous 
structures, such as holes in the chromosphere, to explain the observed 
fluorescence spectra of CO and H$_2$. This is qualitatively consistent 
with the conclusion from the analysis of the infrared CO lines by 
Wiedemann et al. (\cite{wiedemann94}) and Tsuji (\cite{tsuji09}). 
The detection of non-zero CP and DPs presented in Sect.~\ref{sect_obs} 
lends direct support to the presence of inhomogeneous structures. 
It should be noted that 
while the inner CO layer of our models is consistent with the previous
studies, our AMBER observations and modeling have revealed that 
the cool molecular component extends out to 2--3~\RSTAR, much farther than 
considered before.

The chromospheric model of Drake (\cite{drake85}) for Arcturus 
based on the analysis of 
UV emission lines shows that the (electron) temperature already reaches 
$\sim$8000~K at $\sim$1.2~\RSTAR\ and stays approximately constant up to 
$\sim$20~\RSTAR. O'Gorman et al. (\cite{ogorman13}) proposed a modified model, 
in which the electron temperature decreases beyond 2.3~\RSTAR, to reproduce 
the radio fluxes measured at 3--20~cm. In either case, the chromospheric wind 
reaches the terminal velocities of 35--40~\KMS\ already within 
$\sim$2~\RSTAR. On the other hand, our modeling of the AMBER data shows that 
the MOLsphere extends to $\sim$2.6~\RSTAR. This means that the chromospheric 
wind accelerates within the radius of the MOLsphere. 
Nevertheless, neither the spectrum of the CO first overtone lines 
nor the visibilities obtained with AMBER show a signature of this 
outflow in the MOLsphere, which would have been detectable with the 
spectral resolution of 25~\KMS\ of AMBER. 
The analyses of the high-resolution spectra of 
the CO fundamental lines do not show a systematic outflow either 
(Heasley et al. \cite{heasley78}; 
Wiedemann et al. \cite{wiedemann94}; Tsuji \cite{tsuji09}). 
Tsuji (\cite{tsuji88}) inferred that the MOLsphere is quasi-static, 
because even if it does not show a systematic outflow, it is characterized by 
strong turbulence with a turbulent velocity as high as 10~\KMS.

Our AMBER observations and modeling of Arcturus, 
combined with the analysis of the chromospheric emission lines, 
suggest the coexistence of two components within 2--3~\RSTAR\ 
that are distinct not only thermally, but also dynamically: a quasi-static 
cool molecular component, and a steeply accelerating chromospheric wind. 
Figure~\ref{alfboo_schematic} shows a schematic view of the inhomogeneous 
structures of the outer atmosphere of Arcturus. 
The stratification in the chromospheric region is taken from Fig.~6 of 
Ayres et al. (\cite{ayres03}). While their figure shows the chromosphere 
only up to 1.15~\RSTAR, we assumed it to extend to much larger radii 
to represent the chromospheric wind. 
The cool molecular component, which covers most of the surface of 
the star, extends to the same atmospheric heights as the chromospheric 
component (at least to $\sim$2.6~\RSTAR). 
It should be kept in mind that the geometrical thickness of the 
outer CO layer was assumed to be 0.1~\RSTAR\ in our modeling, 
because it cannot be constrained by the current data. 
It is possible or rather realistic 
that the outer layer is geometrically thicker with a continuous density 
distribution. Observations with higher spatial resolution are necessary 
to probe this issue. 

The physical process responsible for the formation of the outer atmosphere 
consisting of the hot chromospheric gas and cool molecular gas is not yet 
understood. 
As discussed in Ohnaka (\cite{ohnaka13b}), the Alfv\'en-wave-driven 
wind models (Suzuki \cite{suzuki07}) 
show highly temporally variable stellar winds consisting of 
hot gas ($10^4$--$10^5$~K) embedded in cool gas (1000--2000~K). The temperatures 
of this cool component are in agreement with the results of our modeling. 
Since the models assume fully ionized plasma, which is not appropriate 
for the outer atmosphere of red giants, it is necessary to compare with 
the observations once the models are further improved. 

The recent 3D simulations of the chromosphere 
of red giants by Wedemeyer et al. (\cite{wedemeyer17}) show that 
the shock waves produced by convective motions 
give rise to filaments of hot gas reaching a temperature of $\sim$5000~K, 
even when magnetic fields are not included. 
The filamentary chromospheric gas is embedded in cooler gas at a 
temperature as low as $\sim$2000~K within an atmospheric height of 
$\sim \!\! 2 \times 10^{11}$~cm, which corresponds to $\sim$1.1~\RSTAR\ 
for the stellar radius of $2.05 \times 10^{12}$~cm predicted by the MARCS 
models for Arcturus. 
The temperature and the atmospheric height of this cool gas 
are comparable to the temperatures and radius of the inner CO layer derived 
from our modeling. 
The models of Wedemeyer et al. (\cite{wedemeyer17}) 
also show that the density of the cool gas at atmospheric 
heights within $\la \! 10^{11}$~cm ($\la$1.05~\RSTAR) is as high as 
$\sim \!\! 10^{-11}$~g~cm$^{-3}$, which broadly agrees with the 
gas density of $1.5 \times 10^{-11}$~g~cm$^{-3}$ of the inner CO layer 
derived in Sect.~\ref{subsect_marcs_wme}.
While the geometrical extension of the 3D models is still too small compared 
to the overall extension of the MOLsphere of 2--3~\RSTAR, it would be 
interesting to compare such theoretical models with the observationally 
derived properties of the MOLsphere, once they are extended to 
larger radii.

As described in Sect.~\ref{subsect_marcs}, RSGs tend to show more pronounced 
emission from the MOLsphere than normal K--M giants. 
Based on high spectral resolution AMBER data, we have derived the 
parameters of the MOLsphere for two RSGs (Betelgeuse and Antares, 
Ohnaka et al. \cite{ohnaka09}, \cite{ohnaka11}, \cite{ohnaka13a}) and 
three normal K--M giants (BK~Vir, Aldebaran, and Arcturus, 
Ohnaka et al. \cite{ohnaka12}, Ohnaka \cite{ohnaka13b}, and this work). 
However, the differences in the effective temperature between two RSGs 
and three red giants do not allow us to discuss possible differences in 
the MOLsphere parameters between RSGs and normal K--M giants. 
We have carried out high spectral resolution AMBER observations 
for a small sample ($\text{about ten}$ stars) of red giants and RSGs 
to study the possible systematic dependence of the MOLsphere properties on 
basic stellar parameters such as luminosity, effective temperature, and 
surface gravity. The results will be reported in a future paper.

\section{Concluding remarks}
\label{sect_concl}

Our high spectral resolution infrared interferometric observations of 
the well-studied red giant Arcturus in the individual 2.3~\micron\ CO lines 
have spatially resolved the molecular outer atmosphere. 
The MARCS photospheric models are too compact to explain the observed data 
in the CO lines, confirming that the extended molecular outer 
atmosphere is not accounted for by current photospheric models. 
The observed spectra and visibilities can be reproduced 
by models in which two additional CO layers are added above the MARCS 
photospheric model. The inner CO layer is geometrically thin 
($\la$0.02~\RSTAR) and is located just above the photosphere, 
at $1.04\pm 0.02$~\RSTAR, with a CO column density of 
$10^{20\pm 0.3}$~\PERSQCM\ 
and a temperature of $1600\pm400$~K. The properties of this layer are in 
agreement with what has been inferred from previous spatially unresolved 
spectroscopic studies in the infrared and ultraviolet. 
The outer CO layer, however, extends out to $2.6\pm 0.2$~\RSTAR , which is much 
larger than considered before, with a CO column density 
of $10^{19\pm 0.15}$~\PERSQCM\ and a temperature of $1800\pm 100$~K.  

Combined with the chromospheric models that are based on the UV emission lines, 
our AMBER observations and modeling 
suggest that the quasi-static cool molecular component extends out to 
2--3~\RSTAR, within which the chromospheric wind steeply accelerates. 
The detection of the non-zero CP and DPs also suggests the presence of 
such inhomogeneous structures.
It is not clear yet whether the cool component remains quasi-static beyond 
2--3~\RSTAR\ or starts to exhibit a systematic outflow at some radius. 
Moreover, it is not clear either to what radius the cool 
molecular component extends, because the CO first overtone lines 
sample a relatively warm region. Spatially resolved observations 
in the CO fundamental lines would be useful to probe the region farther 
out and obtain a more comprehensive picture of the outer atmosphere. 
This will be possible with the second-generation VLTI instrument 
MATISSE (Lopez et al. \cite{lopez14}).

\begin{acknowledgement}
We thank the ESO Paranal team for supporting our AMBER observations. 
We are also grateful to Dieter Schertl and Karl-Heinz Hofmann for their 
help with the AMBER data reduction. 
K.~O. acknowledges the support of the 
Comisi\'on Nacional de Investigaci\'on Cient\'ifica y Tecnol\'ogica
(CONICYT) through the FONDECYT Regular grant 1180066. 
This research made use of the \mbox{SIMBAD} database, 
operated at the CDS, Strasbourg, France, 
and NSO/Kitt Peak FTS data on the Earth's telluric features 
produced by NSF/NOAO.
\end{acknowledgement}

\appendix
\section{Transfer function of the AMBER observations}
\label{appendix_tfplot}

\begin{figure*}
\resizebox{\hsize}{!}{\rotatebox{0}{\includegraphics{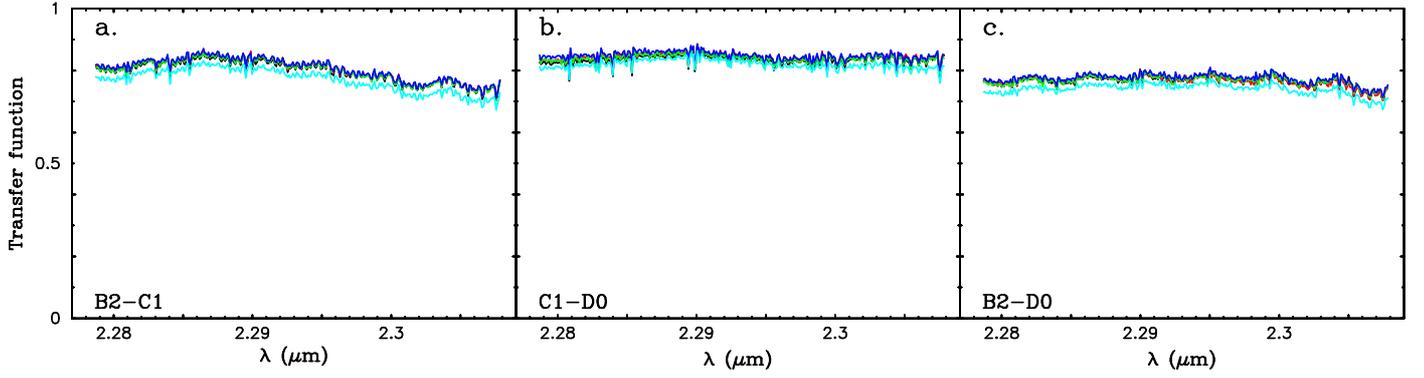}}}
\caption{
Transfer function derived on the night of the observation of Arcturus. 
Panels {\bf a}, {\bf b}, and {\bf c} show the transfer function 
measured at baselines B2-C1, C1-D0, and B2-D0, respectively, 
taking the best 20\% of the frames in terms of the fringe S/N. 
In each panel, 
the transfer functions derived from five AMBER observations of the 
calibrator \alfcenA\ (C1--C5 in Table~\ref{obs_log}) are plotted 
in black (C1), red (C2), green (C3), blue (C4), and light blue (C5). 
The transfer function was stable throughout the night, and the 
five curves therefore nearly overlap. 
}
\label{tfplot}
\end{figure*}

Figure~\ref{tfplot} shows the transfer function 
(i.e., the visibility that would be measured for a point source)
derived from the AMBER observations of the calibrator \alfcenA. 
It was derived by taking the best 20\% of the frames in terms of 
the fringe S/N and used for the calibration of the Arcturus data. 
The figure shows that 
the transfer function values on three baselines are very stable 
throughout the night, thanks to the good and stable atmospheric conditions.

\end{document}